\documentclass[journal]{IEEEtran}
\usepackage[latin1]{inputenc}
\usepackage{times,amsmath}
\usepackage{amssymb}
\usepackage{steinmetz}
\usepackage{pstool}
\usepackage{subfigure}
\let\labelindent\relax
\usepackage{enumitem}
\usepackage{multirow}
\usepackage{enumerate}
\usepackage{graphicx}
\usepackage{mwe}
\usepackage{bigints}
\usepackage{MnSymbol}
\usepackage{stfloats} 
\usepackage[table]{xcolor}
\usepackage[square, comma, sort&compress, numbers]{natbib}
\usepackage{nohyperref}
\usepackage{algorithm,algorithmic}
\usepackage{epstopdf}
\usepackage{comment}
\usepackage{mathtools, cuted}
\usepackage[table]{xcolor}
\usepackage{array}
\definecolor{verylightgray}{rgb}{0.95,0.95,0.95}
\usepackage{multicol} 
\graphicspath{ {Figures/} }
\usepackage{colortbl}

\begin{document}

\title{ Internet of medical things for non-invasive and non-contact dehydration monitoring away from the hospital: state-of-the-art, challenges and prospects }

\author{
\IEEEauthorblockN{
Soumia Siyoucef\IEEEauthorrefmark{1},\IEEEauthorrefmark{2}, Rose Al-Aslani\IEEEauthorrefmark{1}, Mourad Adnane\IEEEauthorrefmark{2}, Muhammad Mahboob Ur Rahman\IEEEauthorrefmark{1}, Taous-Meriem Laleg-Kirati\IEEEauthorrefmark{3}, Tareq Y. Al-Naffouri\IEEEauthorrefmark{1} 
}

\IEEEauthorblockA{\IEEEauthorrefmark{1}Computer, Electrical and Mathematical Sciences and Engineering Division (CEMSE), \\ 
King Abdullah University of Science and Technology, Thuwal 23955, Saudi Arabia} \\
\IEEEauthorblockA{\IEEEauthorrefmark{2}Electronics Engineering Department, LDCCP Lab, Ecole Nationale Polytechnique, Algiers, Algeria} \\
\IEEEauthorblockA{\IEEEauthorrefmark{3}The National Institute for Research in Digital Science and Technology, Paris-Saclay, France\\
\IEEEauthorrefmark{1}\{muhammad.rahman,Tareq.alnaffouri\}@kaust.edu.sa, }
\IEEEauthorrefmark{2}soumia.siyoucef@g.enp.edu.dz,
\IEEEauthorrefmark{3}Taous-Meriem.Laleg@inria.fr
\thanks{The research reported in this publication was supported by funding from King Abdullah University of Science and Technology (KAUST) - KAUST Center of Excellence for Smart Health (KCSH), under award number 5932. } 
}


\maketitle

\begin{abstract}

Dehydration occurs when the body loses more water than it takes in. Mild dehydration can lead to fatigue, cognitive impairments, and physical complications, while severe dehydration can cause life-threatening conditions like heat stroke, kidney damage, and hypovolemic shock. Traditional bio chemistry-based clinical gold standard methods are expensive, time-consuming, and invasive. Thus, there is a pressing need to design novel non-invasive methods that could do in-situ, early and accurate detection of dehydration, which will in turn allow timely intervention. This article presents a methodological review of the literature on a range of innovative internet of medical things-based techniques for dehydration monitoring. We begin by briefly describing the pathophysiology of the dehydration problem, its clinical significance, and current clinical gold-standard methods for assessing hydration level. Subsequently, we critically examine a number of non-invasive and non-contact hydration assessment studies. We also discuss multi-modal sensing methods and assess the impact of dehydration among specific population groups (e.g., elderly, infants, athletes) and on different organs. We also provide a list of existing public and private datasets which make the backbone of machine learning-driven research on dehydration monitoring. Finally, we provide our opinion statement on the challenges and future prospects of non-invasive and non-contact hydration monitoring.

\end{abstract}

\begin{IEEEkeywords}

Dehydration, non-invasive methods, non-contact methods, multi-modal sensing, machine learning, internet of medical things, vulnerable population.  
\end{IEEEkeywords}

\section{Introduction}

About 60\% of the human body is made up of water which highlights the significance of water for several physiological processes and preserving general health \cite{1d}. Humans need at least 1.5 liters of water each day for maintaining optimal body function \cite{2d}. Beyond simple hydration, water helps regulate weight by increasing the body's metabolic rate \cite{Thornton2016-nt}, plays an essential role in building new cell membranes \cite{Cooper2000-xq}, is important for sweating and urination, which aids in removing toxins \cite{Genuis2013-ai}-\cite{Genuis2011-ux}, and aids in temperature regulation and the transport of nutrients \cite{soumia01}.

Dehydration is a prevalent condition that affects 17\% to 28\% of elderly population in the United States \cite{61}. 
Dehydration occurs when water intake does not replace free (hypotonic) water lost due to normal physiological processes, including breathing, urination, perspiration, or other causes such as diarrhea, and vomiting \cite{12d}. 
Remarkably, even a slight decrease in body water, as little as (1-2)\% can significantly impact cognitive and motor function \cite{16d}. 
Thus, dehydration is a major health risk that can lead to a wide range of physiological problems. 

Dehydration could be caused by different reasons, e.g., exposure to heat or sun, hot and humid weather, lack of water intake, intense physical activity, use of diuretics, onset of diarrhea. 
The risk factors associated with dehydration include the following: old age, female gender (due to the higher percentage of body fat), and a Body Mass Index (BMI) $<21$ or $>27$. Individuals with dementia, history of stroke, urinary incontinence, infections, use of steroids, and a decrease in functional independence also increase the risk of dehydration. 
Some other vulnerable populations include sportsmen/athletes, construction workers, infants and children, chronic patients taking diuretics drugs. 
The elderly make a vulnerable group because of decreased body water content, decreased thirst sensation, and more common kidney problems. A research study indicates that dehydration significantly increases the risk of mortality among hospitalized older patients \cite{14d}. Last but not the least, dehydration could exacerbate diseases like childhood obesity due to electrolyte abnormalities \cite{15d}.

Dehydration can be life-threatening when severe, could lead to seizures, and also carries the risk of osmotic cerebral edema if rehydration is carried out rapidly. As for dehydration diagnosis, it is mostly done in a clinical setting whereby a clinician examines the physical symptoms of the patient, such as confusion, lethargy, rapid weight loss, and functional decline. This observation is then combined with the biochemistry-based biomarkers extracted from the blood and urine samples in order to reach a decision. 
In fact, dehydration is one of the most common scenarios a physician/Emergency room of a hospital faces, for incoming patients of all ages (especially children). If diagnosed early, the treatment of dehydration is simple (oral hydration) but if it is detected late then doctors need to administer the intake of necessary fluids through parenteral route (systemic via intravenous and other routes). Thus, seamless, ubiquitous and rapid monitoring of the hydration status of the body (e.g., by means of a small, hand-carried gadget, wearable device.) is the need of the hour. An early intervention following an early detection of severe dehydration could have a life-saving impact.

The gold standard dehydration detection methods include a handful of methods that measure body weight changes, assess urine output and color, conduct blood tests, and use isotopic dilution techniques. Such gold standard methods, despite their accuracy, have their own limitations, e.g., they are invasive and time-consuming, available in clinical setting only,  costly,  not suitable for real-time monitoring, and influenced by confounding factors such as medical conditions. They also may not be sensitive enough to detect mild dehydration. Thus, to address these limitations, researchers and clinicians have started to explore a wide range of alternative non-invasive and non-contact methods. These emerging dehydration monitoring methods aim to provide more accessible, rapid, and accurate dehydration detection, enabling timely interventions to prevent and manage dehydration. Such methods also minimize discomfort for the individuals, and allow for frequent monitoring without interfering with their daily activities, ultimately contributing to the evolution of tomorrow's healthcare systems. In addition, contactless methods are particularly useful during pandemics, as they help to avoid contamination. Contactless methods are also suitable for people with special needs. 


{\it Scope of this survey paper:}
This paper aims to critically examine the recent research works on non-invasive dehydration monitoring, including contactless methods. 



{\it Contributions: } 
There exist a few review articles that aim to summarize the state-of-the-art on dehydration monitoring, but these works either put a narrow focus on a particular disease (e.g., Urolithiasis aka kidney stone disease \cite{tahar2024non}), or specific populations, such as elderly \cite{14d}, athletes \cite{70}. Further, none of the existing review articles have talked about contactless dehydration sensing methods. Similarly, none of the existing review articles provides a systematic discussion about the dehydration-centric datasets.
This methodological review article differs from existing survey articles in a number of ways. First, to the best of our knowledge, this is the most comprehensive and up-to-date survey on dehydration monitoring, to date. This is because we provide a detailed discussion of a wide range of sensing modalities (both contact-based and contactless) and their variants, and provide extensive references. 
To this end, we also provide a detailed comparison table outlining the benefits and limitations of different sensing modalities (as a supplementary document). Our discussion also covers the implications of dehydration on a number of population groups, e.g., children, elderly, chronic patients, athletes. Furthermore, keeping in mind the rise of Artificial Intelligence (AI) methods and noting that data is the key/backbone for success of AI methods, we have also prepared a list of publicly available and private datasets (available upon request) for hydration assessment (see the supplementary document). While making this table, we took care to include key details such as the sensing modalities employed and the devices used for data collection, number of participants... etc. In addition, we provide a timeline of selected existing review papers on non-invasive hydration assessment (2003-2024) from the literature (see Figure \ref{timeline}), highlighting their scope, the sensing modalities discussed, and their key limitations. Further, we devote a full section to non-contact methods which rely upon radio-frequency signals, acoustic signals, optical signals, and camera-based video recordings. Last but not the least, we discuss the challenges and outlook by providing our opinion, position and vision about the way forward, at the end.





{\it Literature review criterion:}
To search for the relevant literature, we scoured various academic research databases, i.e., Scopus, Web of Science, PubMed, IEEE Xplore, ScienceDirect, Google Scholar, Arxiv, and other online resources. Specifically, we searched the relevant literature spanning the years from 2003 to 2024. To retrieve the published articles of interest, we utilized the following keywords in various combinations: dehydration, non-invasive monitoring, non-contact monitoring, blood, urine, osmolality, and sweat. We also searched for dehydration monitoring solutions for diverse populations, including older people, athletes, infants, chronic kidney disease patients, and others. We focused particularly on highly cited and influential works. After gathering the relevant literature, we categorized the works based on the sensing modality used.  

{\it Outline:} 
The rest of this paper is organized as follows:
Firstly, we provide background information on dehydration, covering topics such as the significance of hydration levels as a biomarker, gold standard methods for dehydration monitoring, and the need for novel non-invasive and non-contact monitoring methods. Then, we examine many different sensing modalities used to assess hydration, including fluid based methods, electrodermal activity, electrocardiograph, bioelectrical impedance analysis, acoustic, radio frequency, optical, and thermal sensing. Next, we touch upon multi-modal sensing approaches and discuss hydration assessment in targeted organs and specific populations. Finally, we present our vision for the future of hydration monitoring and conclude the paper. 

\section{A short primer on dehydration}

Dehydration, being a multi-faceted and whole-body phenomenon, has remained a matter of intense debate among the subject specialists, but without any fruitful conclusion. Thus, there is a lack of consensus among the clinicians and physicians on a universal definition of dehydration. Nevertheless, pertinent details such as dehydration types, pathophysiology, gold standard methods are described below. 

\subsection{Dehydration types and pathophysiology}

At any given time, the body of a healthy person is EU-hydrated, i.e., it is neither dehydrated, nor hyperhydrated. But when a person is dehydrated, there are a number of ways to classify and assess the dehydration, i.e., it could be done based on body compartments involved (internal or external), based on severity level, based on water-electrolyte balance, and more.  

{\it Dehydration types (Internal and peripheral):}
Based on the literature on dehydration pathophysiology, we identify two main causes of dehydration: i) internal dehydration that occurs as a consequence of lack of water intake, say, due to fasting, ii) peripheral (skin) dehydration that occurs due to intense physical activity in the sun or heat (e.g., by construction workers and athletes.). Below we discuss the pathophysiology of two types of dehydration, one by one. 

{\it Pathophysiology of internal dehydration:} internal dehydration begins when the body loses more water than it consumes. In this situation, the body triggers a cascade of responses involving specialized sensors in the brain called osmoreceptors, which detect a decrease in water levels and send signals to the hypothalamus to induce thirst. Simultaneously, the kidneys are stimulated to retain water by releasing the antidiuretic hormone, increasing water absorption. Moreover, reduced blood pressure because of dehydration triggers the kidneys to release the renin enzyme, which starts a sequence that ends with the release of the aldosterone hormone, which in turn leads the kidneys to increase water retention. As dehydration progresses, the body shifts its water resources from the outside to the inside, ultimately causing symptoms like dry skin, decreased sweating, and reduced mucous membrane moisture \cite{61}. 

{\it Pathophysiology of peripheral dehydration:} In peripheral dehydration, there is a water deficit in the peripheral tissues due to excess sweating. Here, water moves from the body's interior to the skin's surface for evaporative cooling \cite{b6}. This is because during physical exertion, the body generates heat due to muscle energy transformation. To aid in cooling, Eccrine glands absorb water from the blood and release it onto the skin surface, transferring water from the inside to the outside of the body. This way, the useful electrolytes and ions are also lost through sweat, which negatively impacts the water-electrolyte ratio in the body \cite{b7}. Thus, the water and electrolyte losses need to be replenished by adequate fluid intake to prevent dehydration.

{\it Dehydration types (water-electrolyte ratio-based):}
Another way to examine dehydration phenomenon is to measure the imbalance in water-electrolyte ratio. 
That is, the proportional relationship between free water and Sodium leads to three types of dehydration as follows. 
1) { Hypernatremic or hypertonic dehydration} occurs when more water is lost compared to sodium (e.g., due to infection or exposure to high temperatures). 
2) In { Isonatremic or isotonic dehydration}, there is an equal loss of water and sodium (e.g., due to vomiting and diarrhea). 
3) { Hyponatremic or hypotonic dehydration} occurs when more sodium is lost compared to water (e.g., due to the use of diuretics, typically by the elderly people). 

{\it Dehydration types (severity-based):} In terms of severity,  dehydration could be classified as either mild, or moderate, or severe. Mild dehydration implies less than 50 mL/kg body fluid loss or less than 5\% weight loss; moderate dehydration implies 50 to 100 mL/kg body fluid loss or 5\% to 10\% weight loss; and severe dehydration implies more than 100 mL/kg body fluid loss or more than 10\% weight loss. 



\subsection{Gold standard methods for dehydration assessment}

{\it Blood and urine samples-based methods:}
The gold standard methods begin by collecting blood and urine samples from the patient, and utilize the outcomes of the biochemistry results along with the physical symptoms of the patient, in order to do dehydration diagnosis, dehydration type identification, and severity assessment.
Precisely speaking, the blood sample-based biochemistry analysis provides a number of relevant biomarkers such as serum sodium, plasma osmolality, and more. Among them, plasma/serum osmolality, that measures the body's electrolyte-water balance, is regarded as the most reputed biomarker for dehydration \cite{m1,b4}. On the other hand, the urine sample-based biochemistry analysis provides rich information about a number of dehydration-related biomarkers, e.g., Urine Specific Gravity (USG), urine osmolality, urine color (Armstrong chart), and urine volume \cite{b2}. Another important bio-marker is the Blood Urea Nitrogen (BUN)/creatinine ratio, which is an indirect indicator of volume depletion \cite{66}. In addition to blood and urine based biomarkers, a clinician also examines a number of physical symptoms, such as tongue, skin turgor, eyes, blood pressure, heart rate... etc. Finally, measuring the body mass change over time due to dehydration also helps quantify the extent of dehydration. This is because of the fact that changes in body mass can indicate the volume of water loss (assuming one gram of body mass loss equals one milliliter of water loss) \cite{b3}.
Table \ref{tab:dehydration_biomarkers} outlines the typical values of various blood and urine-based biomarkers for dehydration assessment \cite{tab1,tab11}. 

\begin{table}[h!]
    \centering
    \renewcommand{\arraystretch}{1.3}
    \begin{tabular}{|>{\raggedright\arraybackslash}m{4cm}|>{\raggedright\arraybackslash}m{4cm}|}
        \hline
        \textbf{Biomarker indicator} & \textbf{Threshold} \\
        \hline
        Serum Sodium & 
        1. 135--145 mEq/L (Well-hydrated) \\
        & 2. $>$145 mEq/L (Dehydrated) \\
        \hline
        Urine Specific Gravity (USG) & 
        1. $<$1.010 g/ml(well-hydrated)\\
        &2. 1.010--1.030 g/ml (Slightly dehydrated/ Hypohydration) \\
        & 3. $>$1.030 g/ml(Chronic dehydration) \\
        \hline
        Urine Colour & 
        1. Score of 1--2 (Well-hydrated) \\
        & 2. Score of 3--6 (Slightly dehydrated) \\
        & 3. Score of 7--8 (Dehydrated) \\
        \hline
        Urine Osmolality & 
        1. $<$700 mOsm/kg (Normal) \\
        & 2. $\geq$700 mOsm/kg (Dehydrated) \\
        \hline
        Serum/Plasma Osmolality & 
        1. $\leq$295 mOsm/kg (Normal) \\
        & 2. $>$295 mOsm/kg (Dehydrated) \\
        \hline
    \end{tabular}
    \caption{Some blood and urine-based biomarkers for quantifying dehydration level. (mOsm/kg stands for milliosmoles per kilogram, while mEq/L stands for milliequivalent per liter.)}
    \label{tab:dehydration_biomarkers}
\end{table}

{\it Isotope dilution method:}
Another accurate but less common method is the so-called Isotope dilution method. Under this method, a person ingests an isotope (typically, Deuterium Oxide (D2O), i.e., heavy water) in a known amount. Later, the concentration of the isotope in a bodily fluid (blood or urine) is computed to determine the total body water volume. 

{\it WHO guidelines for dehydration assessment in children:}
World Health Organization (WHO) has provided guidelines to classify the hydration levels of children in the age range of 1 month to 5 years into three categories as follows: no dehydration, some dehydration and severe dehydration \cite{tab2} (see Table \ref{tab:dehydration_scale}). As can be seen, the WHO guidelines are solely based on clinical symptoms of the patient, i.e., alertness, eyes, thirst, skin turgor, that are to be examined by a clinician at the time of presentation.

\begin{table*}[h!]
    \centering
    \renewcommand{\arraystretch}{1.3}
    \begin{tabular}{|>{\centering\arraybackslash}m{3.5cm}|>{\centering\arraybackslash}m{3.5cm}|>{\centering\arraybackslash}m{4cm}|>{\centering\arraybackslash}m{4cm}|}
        \hline
        \textbf{Characteristics} & \textbf{No dehydration} & \textbf{Some dehydration ($>$1 sign)} & \textbf{Severe dehydration ($>$1 sign)} \\
        \hline
        Alertness & Well, alert & Irritable or drowsy & Lethargic or poorly responsive \\
        \hline
        Eyes & Normal & Sunken & Sunken \\
        \hline
        Thirst & Drinks normally & Drinks eagerly & Poor or weak drinking \\
        \hline
        Skin turgor & Goes back quickly & Goes back slowly ($<$2 s) & Returns very slowly ($>$2 s) \\
        \hline
    \end{tabular}
    \caption{WHO scale for dehydration in children aged 1 month--5 years old}
    \label{tab:dehydration_scale}
\end{table*}

It is worth mentioning that there exist other population-specific rules of thumb for dehydration monitoring \cite{who1,who2}.

\begin{figure*}[h!]
\centering
\includegraphics[scale=0.55]{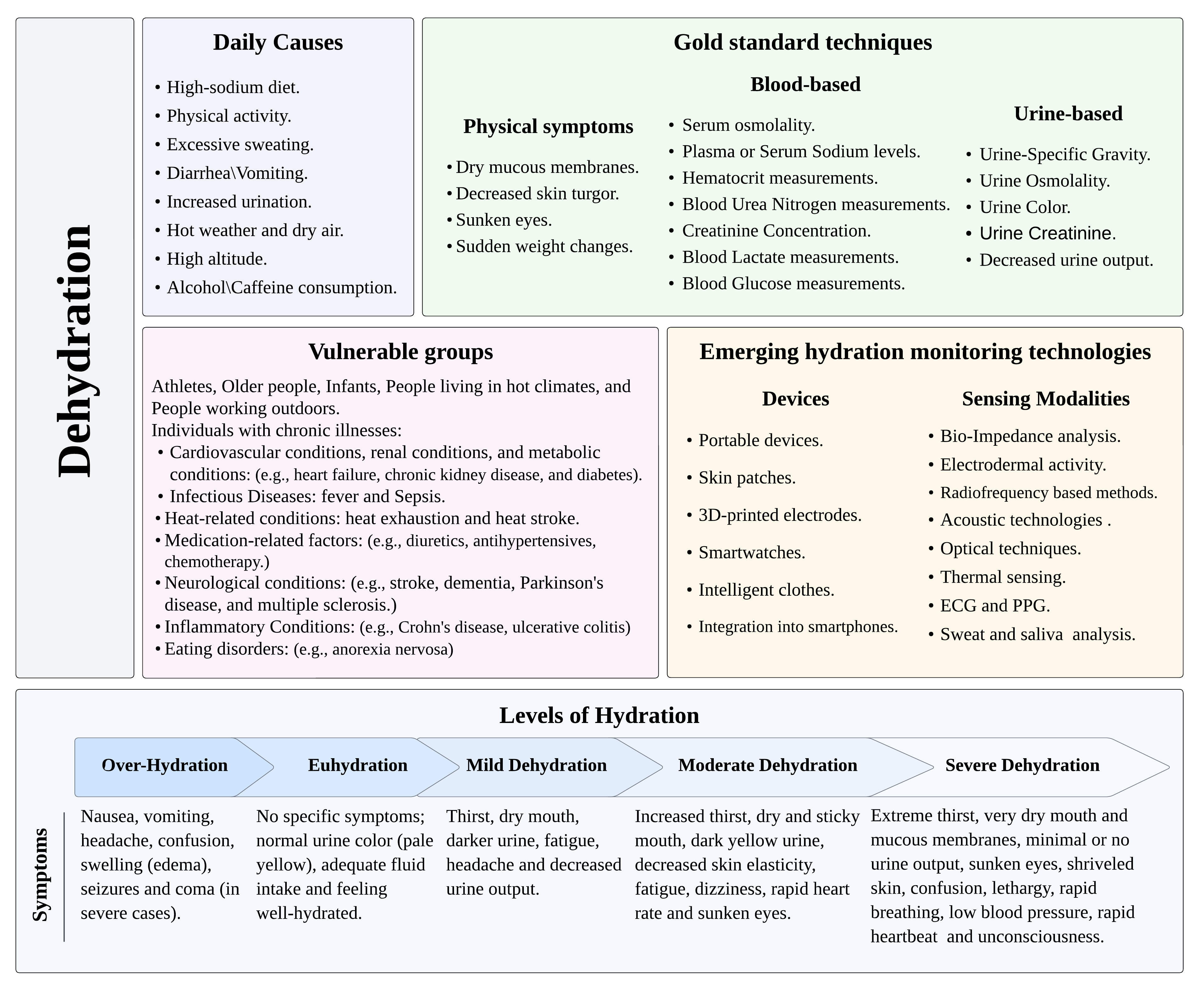}
\caption{A graphical overview/summary of the dehydration monitoring problem}
\label{fig:dehydration_graphical_summary}
\end{figure*}

{\it Pictorial summary of the dehydration problem:}
Fig. \ref{fig:dehydration_graphical_summary} provides a comprehensive graphical summary of the dehydration problem, i.e., its causes, vulnerable populations, gold standard invasive and non-invasive methods, emerging non-invasive and non-contact methods, dehydration levels and physical symptoms caused by them.


\subsection{Emerging non-invasive and non-contact methods for dehydration monitoring }

{\it Limitations of existing gold standard methods:}
Despite their accuracy, existing gold standard methods based upon blood and urine samples have a number of limitations. That is, all these tests are done in a clinical setting by specialized medical personnel, and utilize sophisticated costly equipment for laboratory analysis. Thus, such methods are more suitable for infrequent use than continuous daily monitoring. In other words, such clinical methods are not suitable for frequent self-testing at home by, say, chronic kidney disease patients.
In addition, there are some further caveats of such methods. For example, for the urine based methods, it is imperative to consider past ingestion or medical conditions when relying on urine samples. This is because food, alcohol, caffeine, and illness can all affect urine output \cite{b1}. Similarly, though changes in body mass can indicate the volume of water loss; nevertheless, for body mass changes to accurately reflect variations in water weight, specific conditions must be met. That is, factors such as fluid and food intake, as well as excretion through urine and feces, should be considered \cite{m1}. 


{\it Need for novel non-invasive and non-contact methods: }
Keeping in mind the aforementioned limitations of the gold standard methods, design of novel non-invasive and non-contact methods that allow for in-situ dehydration monitoring is the need of the hour. Such portable, non-invasive, contactless, wearable Internet Of Medical Things (IoMT) devices based solutions (e.g., smartwatches, smart bands, and smart tattoos.) are especially attractive for the people away from hospitals and mega cities. This is exactly what has prompted the researchers to investigate a wide range of sensing modalities for dehydration monitoring, with the aim to make integration of hydration monitoring into daily activities easier. A plethora of studies presented in this paper focus on introducing novel sensor technologies that aim to monitor body moisture levels. While some studies introduce brand-new sensing techniques, others try to improve the existing ones by reducing noise, increasing sensitivity... etc. Below, we parse through the relevant literature, categorized by sensing modality used. 

\begin{figure*}[htb!]
\centering
\includegraphics[width=0.9\linewidth]{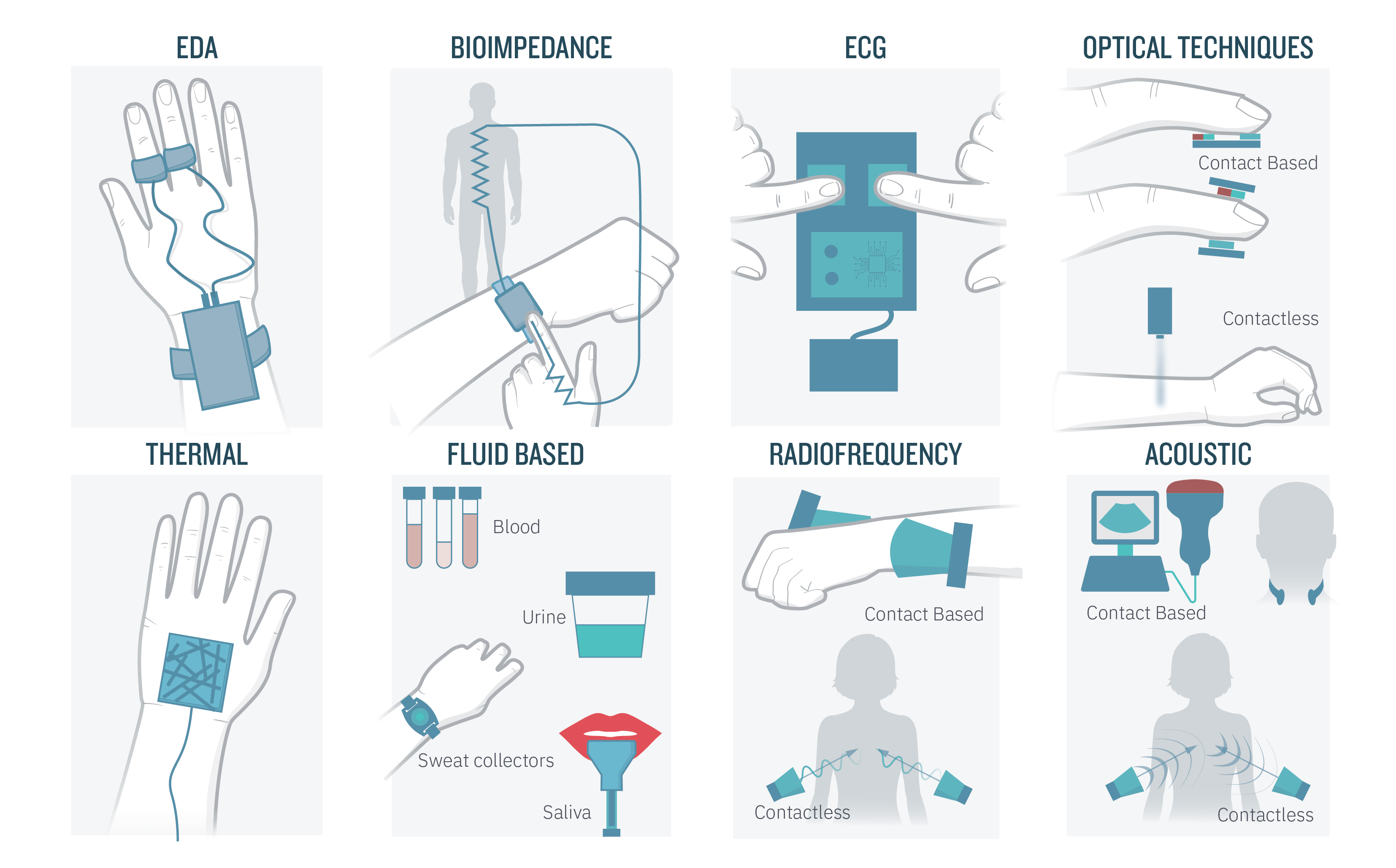}
\caption{Visualization of the eight IoMT sensing modalities that are frequently used for in-situ dehydration monitoring. Among them, radio frequency, optical and acoustic sensing solutions come with contactless variants, while the remaining are all contact-based. }
\label{fig:non-contact-non-invasive-methods}
\end{figure*}

{\it Illustration of non-invasive and non-contact dehydration monitoring methods:}
Fig. \ref{fig:non-contact-non-invasive-methods} illustrates the eight contact-based and contactless sensing modalities for dehydration monitoring reported in the published literature that are discussed in this survey article.

\begin{figure}[htb!]
\centering
\includegraphics[width=0.9\linewidth]{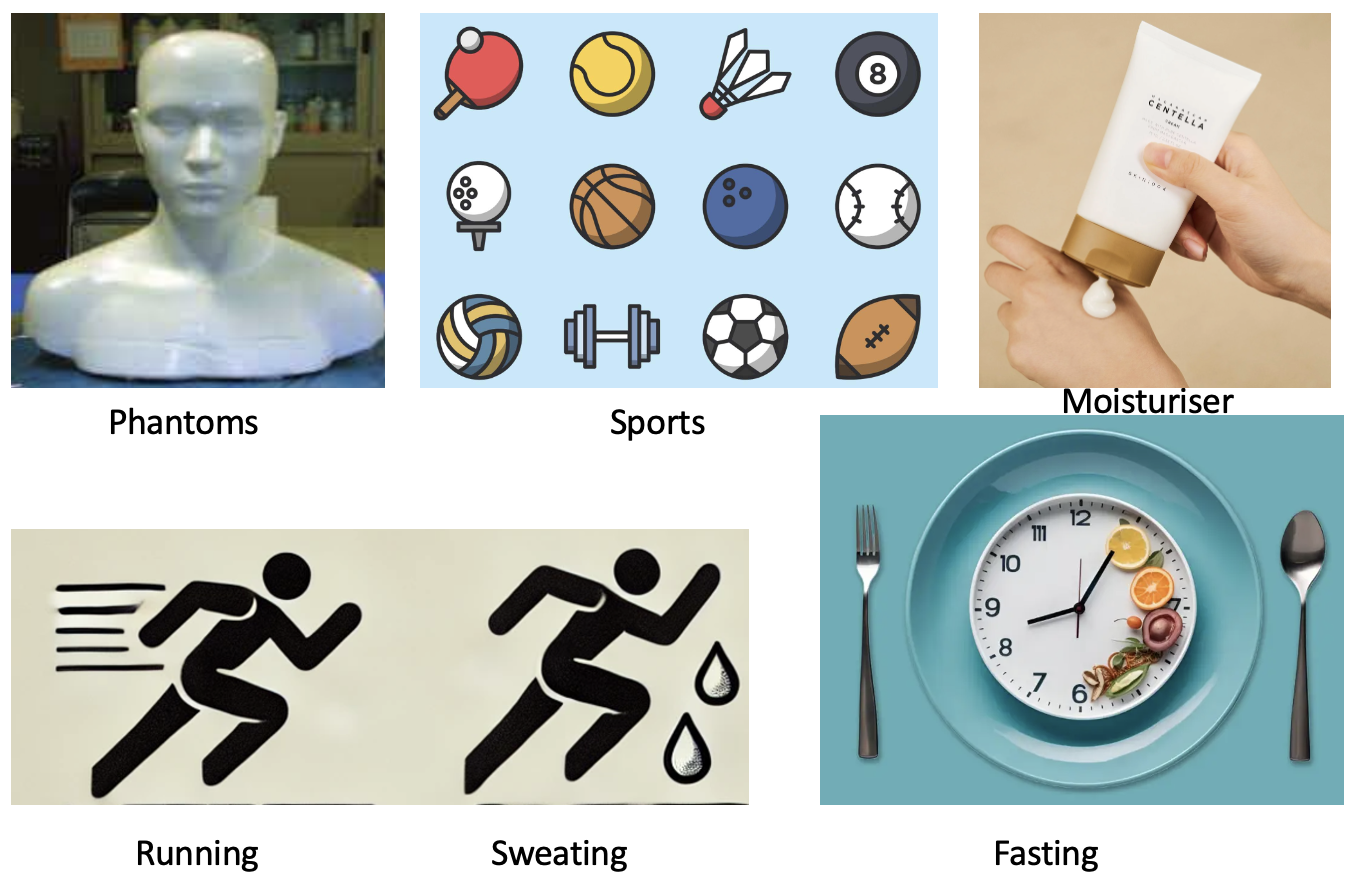}
\caption{Data labeling strategy by IoMT-based non-invasive \& non-biochemistry methods:
All the non-biochemistry methods rely upon exercises such as sports to produce sweat, fasting, gel and moisturizers, and phantoms for obtaining labels for dehydration data being collected. Thus, all these methods are good at dehydration classification only, and are not capable of doing regression, say, estimation of plasma osmolality. }
\label{fig:GT}
\end{figure}

\begin{figure*}[h!]
\centering
\includegraphics[scale=0.41]{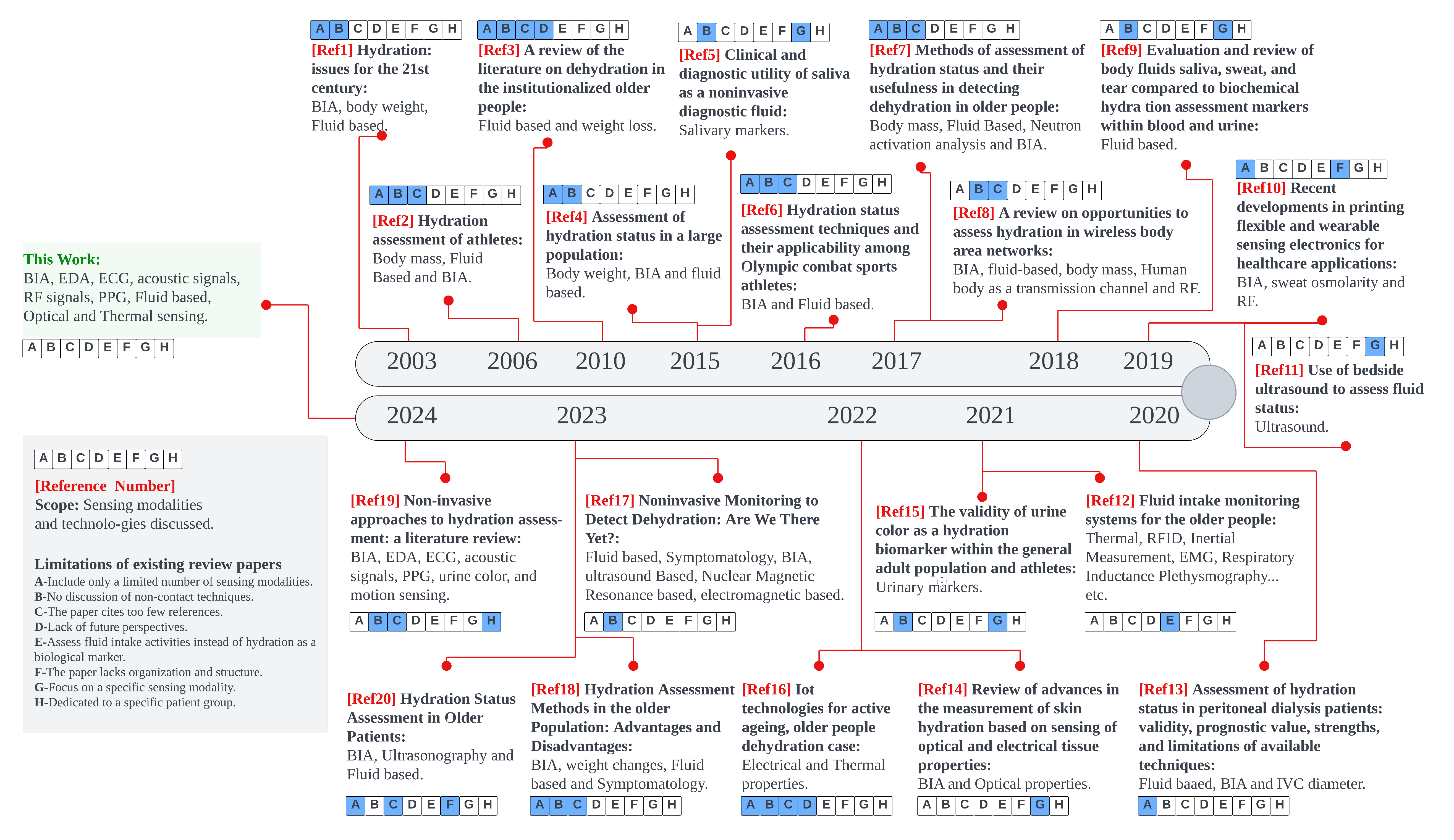}
\caption{Timeline of selected existing review papers on non-invasive hydration assessment (2003-2024) from the literature,  highlighting the scope, sensing modalities discussed, and key limitations. Ref1:\cite{p15} ,Ref2:\cite{70}, Ref3:\cite{p11}, Ref4:\cite{p5}, Ref5:\cite{p12}, Ref6:\cite{p6}, Ref7:\cite{60}, Ref8:\cite{p16}, Ref9:\cite{p14}, Ref10:\cite{p3}, Ref11:\cite{p9}, Ref12:\cite{p2}, Ref13:\cite{p17}, Ref14:\cite{p10}, Ref15:\cite{p13}, Ref16:\cite{p1}, Ref17:\cite{p4}, Ref18:\cite{p7}, Ref19:\cite{tahar2024non}, Ref20:\cite{p8}.}
\label{timeline}
\end{figure*}

{\it Data labeling strategy by IoMT-based non-invasive \& non-biochemistry methods:}
We note that all the methods are non-biochemistry methods (except fluid-based methods). Thus, they rely upon exercises such as sports to produce sweat, fasting, gel and moisturizers, and phantoms for obtaining labels for dehydration data being collected. Thus, all these methods are good at dehydration classification, and are not capable of doing regression, say, estimation of plasma osmolality (see Figure \ref{fig:GT}).

\section{Fluid-based Methods}
Non-invasive fluid-based methods measure biomarkers in bodily fluids (sweat, saliva, urine, tears) that are obtained without penetrating body tissues in an invasive manner. Sweat, secreted by sweat glands, contains water and electrolytes and aids in thermo-regulation. Saliva, produced by salivary glands to moisturize the mouth, consists mainly of water and enzymes. Urine, primarily composed of water and dissolved solutes, is produced by the kidneys for waste elimination. A number of biomarkers could be obtained by doing biochemistry analysis of sweat, urine, saliva, and urine, e.g., osmolality, ion concentrations, serum creatinine, USG, which could provide fine-grained insights into hydration status. During dehydration, these bodily fluids get directly affected due to their high water content, making their analysis valuable for assessing hydration status. This section aims to summarize existing research on non-invasive salivary, sweat, and urine-based methods.

\subsection{Sweat-Based Methods}
The sweat-based methods aim to measure the concentration of electrolytes, such as chlorine, sodium, and potassium in sweat sample. 
Tracking ion concentrations in sweat provides a direct indication of hydration levels in the body. It has been observed that the concentration of electrolytes tends to increase in dehydrated subjects \cite{71}. Additionally, total body water loss has been found to significantly correlate with electrolyte concentrations in sweat, particularly chloride \cite{7320006}. 
Sweat analysis can be performed using electrochemical reactions with ion-selective components \cite{R1}, \cite{41}, through pH assessment \cite{46}, or using crown ethers that are sensitive to sodium ions \cite{50}. Nanoparticles can also be used to track sodium concentration; they produce observable color changes--from red to purple to gray--when their colloidal stability changes. These color changes can be easily detected with am ultraviolet-visible spectrometer or even by the naked eye \cite{52}. \\
\indent Given the ionic nature of electrolytes, the electrical properties of sweat can serve as indicators of ion concentrations \cite{R2}. For example, one could measure conductivity by applying an alternating current to the sweat sample, whereby a higher conductivity corresponds to a higher concentration of ions \cite{R4}. Studies have shown that conductivity of nano-structured copper oxide films changes with varying ion levels, which in turn reflects the body's hydration status \cite{R7}. Additionally, one could utilize (the reflected) radio frequency signals to measure the permittivity of sweat sample; the variations in the reflected signal correlate with the electrical properties of sweat, enabling contactless hydration monitoring \cite{R5}. Sweat volume can also be translated into capacitive values, allowing for the assessment of hydration levels by tracking changes in capacitance instead \cite{R8}.

\subsection{Salivary Biomarkers}
Since saliva is primarily composed of water, its effectiveness as an indicator of hydration has been clinically validated by comparing it to serum osmolality, a standard measure of hydration status \cite{R_saliva_1}. Several salivary markers have been identified as reliable indicators of hydration levels, including electrolyte and hormone concentrations \cite{7539317}, total protein concentration \cite{74}, and saliva osmolality, which has been tested using both real and artificial saliva samples \cite{44}. To provide one concrete example, Stewart et al. have developed embedded piezoresistive microcantilever sensors that can measure osmolality changes in body fluids, including saliva. When the sensing material contacts analyte molecules, it undergoes a volumetric change, creating strain on the microcantilever and altering its resistance--which constitutes a novel approach to hydration assessment \cite{44}.

\subsection{Urine-based Methods}
Urine, as the primary excretory fluid, provides valuable insights about hydration status of a person. Reduced hydration levels lead to higher concentrations of solutes in the urine, such as glucose \cite{R_Urine_3} or ions like potassium \cite{33}.
Urine color, USG, and osmolality are key metrics for assessing hydration levels using urine \cite{R_Urine_1, R_Urine_2}. For example, Ersoy et al. evaluated the hydration status of 26 young male soccer players using urine strips, demonstrating that urine analysis is a reliable and convenient method for athletes to monitor their hydration \cite{R_Urine_1}. In a different study focusing on children, Faidah et al. identified dehydration in subjects based on established USG thresholds and found a strong correlation between USG and salivary osmolality, highlighting the relationship between various bodily fluids and their response to dehydration \cite{R_Urine_5}. 

\subsection{Discussion}
Saliva osmolality, urine color, urine osmolality and USG are reliable biomarkers to determine hydration status of a person. Further, sweat analysis allows continuous dehydration monitoring during physical activities. Thus, fluid-based techniques could serve as non-invasive dehydration detection methods, alleviating the need for expensive and inconvenient invasive methods, which makes them amenable to widespread adoption. However, taking sweat, saliva, or urine samples might not always be convenient or possible. For example, it is typically not possible to collect sweat samples, unless one engages in a physical activity. In addition, collecting fluid samples may require time and effort; subjects have to respect hygienic practices to avoid sample contamination as any interference of the fluid with other substances may lead to a perturbation in values of fluid-based biomarkers leading to false diagnosis. Furthermore, some people may feel uncomfortable providing fluid samples, especially at places where privacy is limited. Thus, fluid-based methods seem appropriate for select population groups only, e.g., sportsmen, construction works outdoors. Furthermore, it is worth mentioning that the use of osmometry to assess dehydration has many limitations \cite{49}, mainly due to the variety in the composition of urine, blood, and saliva among the subjects. Many factors, such as protein metabolism, dietary intake and timing, can influence these body fluids, which in turn affects urine osmolality, while oral artifacts and salivary flow may impact saliva osmolality. Despite these inconveniences, fluid-based analysis could still be employed in clinical interventions to provide quick, easy, and reliable hydration assessments. Cheuvront et al. have introduced practical measures and empirical thresholds to enhance the accuracy of fluid-based methods \cite{49}.

\section{BIA-based methods}
Bioelectrical Impedance Analysis (BIA) is a technique that measures the characteristic impedance of various biological components (e.g. tissues, fats, bones) at many different frequencies. BIA technique has traditionally been used for body composition analysis for conditions like obesity, malnutrition, and fluid retention issues (e.g., lymphedema or heart failure). Nevertheless, multi-frequency BIA technique could reliably obtain very fine-grained details about hydration too. This is done by passing very small amounts of electric current through the body at many different frequencies (in the range 1 KHz - 1 MHz, for a high-end BIA device). This helps obtain an estimate of intracellular water (through high frequencies), extra cellular water (through low frequencies) and total body water. 

Over the years, many innovative BIA sensor designs have been developed to ensure reliable and stable contact with the skin for accurate measurements. These designs include 3D-printed electrodes \cite{5}, textile-based sensors \cite{4}, ultra-thin wearable tattoo sensors \cite{40}, and fully integrated sensors embedded into clothing \cite{54}. Shanshan et al. introduced a capacitive sensor composed of two parallel electrodes made from silver nanowires embedded in a polydimethylsiloxane matrix, providing both stretchability and flexibility to maintain conformal skin contact while eliminating interference from external humidity \cite{26}. Huang et al. proposed an ultra-thin, stretchable sensor system with a differential configuration array of miniaturized impedance measurement electrodes to mitigate common-mode interference, demonstrating superior precision, accuracy, and repeatability compared to commercial devices \cite{28}.

Measured bioimpedance values can be influenced by various factors, such as operating frequencies, sensor spacing, humidity, sweat, and skin temperature \cite{45}. However, these limitations can be mitigated through the use of well-designed materials and advanced signal processing and AI algorithms. 
For example, nanomesh electrodes have been designed which have low sensitivity to water vapor, which in turn helps minimize the impact of transepidermal water loss on measurement accuracy \cite{30}. Additionally, the bias effect of temperature on BIA measurements can be reduced by up to 71\% using Machine Learning (ML)-based correction models \cite{7182268}.

Another related approach that has shown promise involves incorporating a body part into an LC resonant circuit, where a small capacitive electrode is placed in direct contact with the skin which completes the LC circuit. When hydration levels change, the body's added capacitance causes shifts in the resonance frequency. Measuring these frequency shifts, rather than the impedance itself, offers greater sensitivity and accuracy because even small changes in impedance result in significant frequency shifts, simplifying analysis and enabling precise hydration assessment \cite{53}. This method has achieved 92\% accuracy in classifying sportspeople as either hydrated or dehydrated, and 87\% accuracy in classifying the hydration level of fasting individuals \cite{siyoucef2024} on a scale of 1 to 4.
Taking this idea forward, \cite{b8} utilizes the smartphone capacitive touchscreen panels to measure the skin hydration level, thus, eliminating the need for additional sensors and integrating hydration monitoring feature into everyday routine.

Last but not the least, the fact that the body's bioimpedance decreases with increasing hydration levels \cite{36}, allows researchers to study another related phenomenon as well, i.e., it is possible to capture different phases of skin response, in addition to hydration assessment. For example, by infusing an electrolyte solution into superficial voids and analyzing the BIA data, various stages of skin response can be observed \cite{34}. Initially, there is a rapid phase where impedance increases dramatically due to the presence of electrolytes in the voids, followed by a slower phase associated with changes in the skin barrier \cite{34}.

\subsection{ Discussion}
The measurement of characteristic impedance of human tissues is a promising method for hydration assessment. Advantages include provision of real-time results, non-invasiveness, and ease of use. We could also integrate it into wearable devices such as smartwatches, smartbands... etc. However, we note that this sensing modality is based on the contact of the electrodes with the skin, which requires specific stretchable, flexible, and adequate sensor designs. In addition, we must consider the effects of external factors on the readings. That is, the subjects must be in particular postures and stay still during measurements, as otherwise there will be errors and artifacts in the measurements. Furthermore, placing the sensors in body regions where sweat is less present is essential since this could affect the measurements and lead to wrong indications of hydration levels. 

\section{EDA-based methods}
Electrodermal Activity (EDA) sensing measures the skin's electrical conductance which is used as a proxy for assessment of skin hydration. 
Electrode placement and body posture are crucial factors affecting EDA signals, as these signals are sensitive to body stimuli and movements. According to \cite{62}, the palms of the non-dominant hand are among the most suitable areas for electrode placement. The impact of posture on EDA signal accuracy was highlighted in \cite{39}, where 89.7\% accuracy was achieved for sitting postures, while 85.2\% accuracy was reported for standing postures. This study transformed the EDA data into 2D using short-time Fourier transform and passed it to a convolutional neural network.

Since EDA sensors interact with the skin, hence they can be seamlessly integrated into wristbands and smartwatches, enabling real-time tracking of one's hydration level via smartphone applications. This capability is highlighted in \cite{S2023SmartphoneBS}, which utilized EDA data and implemented a dynamic time warping algorithm for dehydration monitoring. Similarly, Kulkarni et al. also developed a smartphone application that employed convolutional neural networks, achieving 84.5\% accuracy in classifying the hydration status of the subjects \cite{kulkarni2021non}.

\subsection{Discussion}
EDA sensor comes embedded into wearable devices, e.g., smartwatch. Thus, it could be utilized for real-time monitoring of one's skin hydration level. However, accuracy of the EDA sensing based hydration level classifiers can be affected by sweat and the stimulation of the nervous system due various emotional states (e.g., stress, anxiety.). Further, EDA sensing is also influenced by the subject's posture, and sudden movement, making it necessary to remove all such artifacts before the data is passed to ML classifiers. Therefore, specific protocols and precautions related to postures, gestures, and the sensor's placement must be followed while collecting EDA data. Otherwise, elimination of these stimulus effects and artifacts in order to enhance the quality of EDA data is a must for EDA-based dehydration monitoring solutions.

\section{ECG-based methods}
The ElectroCardioGram (ECG) is the de-facto gold-standard method to record the heart's electrical activity through a number of electrodes placed on the chest. It is a crucial tool in cardiology for evaluating heart rhythm and diagnosing a number of pathologies such as arrhythmia, myocardial infarction, congestive heart failure, and more. Yet, to the reader's surprise, there have been some attempts to utilize the ECG signals to infer body hydration levels.

ECG parameters have been found to correlate with blood and urine-based biomarkers at different stages of hydration, demonstrating their utility in detecting dehydration \cite{ecg2}, and even during physical activity \cite{59}. Kamran et al. recorded ECG data during various movements to assess hydration states, achieving an AUROC (Area Under the Receiver Operating Characteristic Curve)---a metric used to evaluate the performance of ML models---of 0.89 for classifying hydration levels during supine-to-stand and toe-touch movements. Additionally, Severeyn et al. distinguished between different phases of hydration in 17 male athletes by analyzing various parameters extracted from the ECG signal, such as scaled exponents and approximate entropy \cite{ecg1}.

\cite{mit-bih,mit-bih2,ecg0} have utilized the MIT-BIH PhysioNet database, and reported a sensitivity of 96.3\% using both temporal and spectral features of the ECG signal with an SVM classifier. They identified the RR interval--the time interval between two successive peaks on the ECG waveform--as a crucial feature for differentiating between different hydration levels \cite{alvarez2019machine}. Furthermore, studies have shown that the amplitude of the ECG signal also changes with fluctuations in body hydration \cite{7133644}. The QRS interval was also found to contribute positively to the classification of hydration states, achieving up to 98.73\% accuracy \cite{S-L-2024}. 

The flip side is that there exists a negative example as well, in the ECG research literature on dehydration monitoring. In \cite{ecg3}, basic cardiac electrophysiology was evaluated across a wide range of dehydration levels, i.e., (1-7)\% of total body weight loss, and no evidence was found that cardiac electrophysiology was impaired.

\subsection{Discussion}
ECG technology holds promise for monitoring hydration levels; it is widely available and accessible, offering non-invasive and potentially real-time assessment. Studies demonstrate an existing correlation between several markers extracted from the ECG and the hydration state of a person. However, negative results have also been reported; thus, interpreting the ECG signal to extract hydration information might still be challenging when tested on diverse populations. Also, there is a need to improve the sensitivity and specificity, when using ECG to monitor hydration levels. Last but not the least, ECG alone may not provide a complete hydration assessment; i.e., one may need to fuse data from additional modalities to achieve satisfactory results.

\section{Acoustic-based methods}
Acoustic signals are pressure oscillations that propagate through a medium. Changes in the frequency or amplitude of these pressure signals are analyzed by ultrasound machines, ultrasonography machines, and ML methods for various kinds of health sensing. 

Much of the research efforts to design acoustic sensing-based dehydration monitoring methods rely upon the measurement of diameters of two important blood vessels, i.e., the Inferior Vena Cava (IVC) and aorta (AO). IVC is a large vein that carries deoxygenated blood from the lower body back to the heart, while the AO is the largest artery in the body, responsible for transporting oxygenated blood from the heart to the rest of the body. These two blood vessels play crucial roles in the circulatory system. 

For the sake of dehydration monitoring, the diameters of IVC and AO are assessed using ultrasonography.
Since water is a major constituent in blood, the size of these major vessels is influenced by the body hydration levels. 
There are studies that utilize the IVC diameter to assess hydration level in older patients \cite{25}. Additionally, the ratio of IVC diameter to AO diameter has demonstrated a sensitivity of 82\%, specificity of 91\%, and accuracy of 87\% in predicting significant dehydration in infants \cite{14}. This method has been shown to outperform the WHO's dehydration scale \cite{13}. 

Despite the promise of IVC and AO diameters as biomarkers of one's hydration level, there are several factors that can affect the two diameters. For example, a study has revealed that, in addition to dehydration, C-reactive protein levels and body temperature also significantly influence IVC and AO diameters \cite{15}. Further, blood vessels could also dilate due to other reasons, e.g., during exercise, blood vessels dilate to supply adequate oxygen and nutrients to tissues. One way to mitigate the impact of such factors is by establishing specific cutoff thresholds \cite{25}, allowing for a more accurate dehydration assessment. 

Yet another biomarker is the velocity of a sound wave through a medium that is influenced by the medium's density, elasticity, and moisture content. Therefore, ultrasound velocity can be used to assess hydration levels by placing ultrasonic devices around muscles, such as the calf muscles, to transmit signals and record their propagation speed \cite{11}. Tests on young and older individuals, athletes, and simulated tissues, along with validation against parameters like plasma osmolality, USG, and body mass, have demonstrated the effectiveness of this method \cite{11, 16}. However, researchers have noted significant inter-subject variability and emphasized the importance of monitoring changes in velocity rather than relying solely on absolute values \cite{11}.

In addition to dehydration monitoring in a reactive manner, some other innovative and proactive approaches do exist. For example, \cite{7759295} presents a wearable system based on a microphone placed near the throat, to detect sounds due to drinking activities. Their proposed solution utilizes ML for inference, and comes with a smartphone application, providing users with a daily summary of their water consumption with an accuracy exceeding 90\%.


\subsection{Discussion}
Acoustic technologies are entering the field of hydration assessment as an essential tool. They provide information about muscle moisture levels by tracking the signals that travel through them, thus helping in the non-invasive identification of dehydration. Also, studies using ultrasound sonography to measure the diameter of the inferior IVC and AO have been more accurate in estimating hydration status in patients than other clinical methods. The IVC/AO ratio proves to be an outstanding indicator of hydration states, according to the results of several studies. Nevertheless, variability in the signal and other interfering factors may still be challenging; there is a need to consider variability in the signal instead of direct readings and the determination of specific cutoffs tailored to, for example, age and weight to help in the more accurate estimations. Moreover, the application of ultrasound sonography is a machine-specific process that requires trained staff to interpret the results. It is a clinical intervention that is hard to offer daily for monitoring hydration, therefore the challenge is to miniaturize the devices and to make them more affordable so that they are suitable for everyday use. 

\section{RF-Based methods} 
Radio frequency (RF) waves are electromagnetic waves with frequencies ranging from 3 kHz to 300 GHz. Microwaves are the RF waves that have frequencies between 300 MHz and 300 GHz. 
RF-based methods monitor moisture levels by assessing changes in the dielectric properties of human tissues or skin, such as the earlobe \cite{3}, forearm \cite{11}, or calf \cite{11}. RF waves, particularly at high frequencies, interact with molecules like water. This fact enables researchers to do hydration monitoring using RF-IDentification sensors embedded in textiles that detect dehydration once clothes are 58\% saturated with body sweat \cite{42}. The presence of free water molecules in a sample could also be detected by analyzing the reflected microwave signal using methods like spectral analysis \cite{51}, S-parameter measurements \cite{7}, and time-domain reflectometry \cite{31}. Yet another approach is to utilize the principle of electromagnetic resonance for efficient hydration monitoring \cite{63}. Based on this principle, researchers have developed BioMindR device which has  been tested by both in-vivo and in-vitro experiments, and has shown distinct trends for hydrated and dehydrated states \cite{43}.

In addition, dielectric permittivity has been shown in numerous experiments to be a strong indicator of hydration levels \cite{55}, correlating strongly with body weight changes \cite{57}. Schiavoni et al. introduced a predictive equation for skin hydration status based on tissue permittivity using microwave reflectometry \cite{1, 2}. Various tests, including those conducted on simulated body parts \cite{9}, athletes \cite{10}, fasting subjects \cite{57}, liquid and solid phantoms \cite{7, 8}, and polyester fiber \cite{8}, have confirmed that permittivity is indeed a promising parameter for assessing hydration levels.

\subsection{Discussion}
Researchers have explored many innovative methods to assess hydration using RF-based technologies. Techniques such as time-domain reflectometry, ultra-wideband, S-parameters analysis, dielectric permittivity analysis... etc., have proved their efficiency for monitoring hydration levels, enabling quick and continuous monitoring suitable for real-time applications. RF-based techniques have demonstrated their ability to detect subtle changes in moisture level by assessing the dielectric permittivity of tissues. However, the downside is that this sensing modality requires specific, costly equipment that should be installed with care and needs to be calibrated over time. Furthermore, it is need of the hour to integrate this technology into daily-use devices such as smartphones, wearables, WiFi routers... etc. Therefore, future work must focus on the miniaturization and portability of the devices, making them more affordable and plug-and-play. Another important issue one should consider is to design solutions to fully or partially eliminate radio interference for more accurate estimations.

\section{Optical-Based Methods}
Optical sensing techniques illuminate the object of interest with light of different wavelengths and intensities, and subsequently analyze reflected light waves that have undergone reflection, scattering, absorption, and refraction. 

A well-celebrated optical technique is PhotoPlethysmoGraphy (PPG)---a non-invasive method that can detect variations in blood volume within the microvascular tissue bed. PPG sensors work by illuminating a section of the skin, such as the forehead, ears, or fingers, and detecting the light that is reflected, absorbed, or scattered by the network of blood vessels \cite{reljin2018automatic}. Changes in blood flow within the vessels cause variations in light absorption, providing direct information about specific physiological characteristics such as heart rate, blood pressure, and stroke volume. In addition, the PPG measurements can indirectly offer insights into hydration status. A classification accuracy of up to 70\% has been reported for distinguishing hydrated subjects from dehydrated subjects based on PPG data \cite{ppg2}. 
Another interesting work records short videos of the fingertips using off-the-shelf smartphones, extracts PPG data, and utilizes ML tools to differentiate between four levels of hydration with an accuracy greater than 90\% \cite{alaslani2024you}.

Another widespread method is spectroscopy, which has been explored for dehydration monitoring. One such method utilizes spatially resolved diffuse reflectance spectroscopy and tests it under various stressors, demonstrating precise hydration assessments and revealing different stages of dehydration and rehydration rapidly \cite{75}. During thermal stress, this technique detected a continuous decrease in hydration levels, with no recovery observed until 20 minutes after heat exposure ceased. It was also effective in patients undergoing diuretic therapy for edema syndrome, showing decreased hydration levels due to reduced dermal thickness. Additionally, \cite{20} utilizes infrared spectrometry, obtains a set of features, e.g., capillary refill time, skin recoil time, skin temperature profile, and skin tissue hydration, and estimates hydration levels with 90\% sensitivity and 78\% specificity.

Other than the classical problem of monitoring of body dehydration level, optical methods have also been used to monitor abnormal fluid build up in the body, and for other related problems. \cite{ppg1} utilizes PPG and spectroscopy tools together, to identify congestive heart failure patients, who often experience swelling of the legs and ankles due to their heart's inability to pump blood effectively, causing fluid to accumulate in various parts of the body. In the near-infrared region, distinct scatter graph patterns and second-derivative spectra can differentiate between individuals who use moisturizers frequently and those who do not \cite{23}. 

\subsection{Discussion}
To date, a number of optical techniques have been proposed and tested, including near-infrared spectroscopy, diffuse reflectance spectroscopy, smartphone-acquired PPG, skin temperature profiles, PPG... etc. Optical sensing demonstrated fast response, strong robustness against applied stressors, and gave correct indications about skin properties. However, this sensing modality is prone to environmental factors like dust and ambient lighting, which requires an initial setup and calibration that can be complex. Moreover, validation against the gold-standard methods is needed to prove the evidence of the results and their capability in clinical intervention use. Another major challenge is inter-subject variability, which is influenced by factors such as skin tone, skin thickness, aging, and temperature, as well as the inability to detect subtle changes in degrees of dehydration \cite{ppg3}.

\section{Thermal Sensing-Based Methods}
Thermal sensing detects and measures temperature variations using temperature-sensitive sensors. Thermal transport properties, such as thermal conductivity, specific heat capacity, and thermal diffusivity, correlate positively with hydration levels \cite{73}. These properties, representing the heat transfer characteristics through tissues, have been evaluated in polymers, porcine skin, and human subjects, sometimes outperforming clinical methods in assessing skin hydration \cite{24}. Thermal sensing is frequently used in dermatology, whereby thermography can identify localized temperature changes on the skin's surface, which may indicate inflammation or infection.\\
\indent In all these investigations, skin hydration level is obtained as a bye-product. For example, Shin et al. design a skin hydration sensor based on the transient plane source method in order to determine thermal properties of the skin. Their device uses two resistive heaters and two temperature sensors, one near the heaters and the other at a distance. By comparing the temperature difference between these sensors, the skin's thermal conductivity can be calculated, providing information about the skin's hydration state \cite{72}. While acquiring thermal data, it is important to consider factors like local blood flow dynamics, skin structure, and ambient temperature, which may affect these measurements \cite{73}. The technology proposed by Shin et al. indeed takes into account the issue of fluctuating ambient temperatures. When tested on 200 patients in dermatology clinics, the device shows a 135\% increase in sensitivity and a 36\% improvement in repeatability compared to previous sensors of the same general type \cite{72}.\\
\indent Thermal sensing also finds its use in sports, whereby monitoring skin temperature can help assess hydration levels, as well as few other details that are of interest to the physician, e.g., muscle recovery, signs of overexertion or injury. 

\subsection{Discussion}
Due to the complexity of understanding skin properties in general, using thermal transport properties alone may not provide complete information about skin hydration. Further, the performance of thermal sensing frequently gets degraded due to interfering heat sources. Eliminating external heat sources or the impact of ambient temperature on thermal measurements obtained can be challenging. Therefore, combining thermal sensing with other types of modalities may aid in enhancing the estimation of skin hydration levels and lead to a better understanding of the tissue content. 

\section{Non-contact methods}

In post covid-19 era, governments, healthcare facilities have expressed a great interest in tool that could allow rapid remote monitoring of patients at a mass level. This has prompted the researchers to design a range of contactless methods for various health problems, e.g., covid19 diagnosis, respiratory performance assessment, vitals monitoring, fall detection, and more. The contactless methods rely upon radio signals (WiFi and radar signals), acoustic signals, optical signals, and camera-based video recordings to get the raw data which is ingested by the machine learning algoritms to produce the required inference.  

{\it Non-contact optical methods:}
Non-contact variants of optical methods for dehydration monitoring do exist. One non-contact optical method, validated against the Corneometer CM 825, involves using a laser to illuminate the skin while applying periodic vibrations with a controlled source. This technique tracks back-reflected secondary speckle patterns to estimate various biological markers, including hydration \cite{21}.

{\it Non-contact RF methods:}
Going beyond contact-based sensing, RF technologies also enable non-contact hydration monitoring. By capturing RF signals reflected from or passing through body parts such as upper chest and hand, it is possible to do remote/contactless hydration assessment with an accuracy of up to 93\% in classifying hydrated and dehydrated subjects \cite{hasanhydrationpaper}. Such methods find their utility at public places and events, e.g., malls, marathons, construction sites, for rapid, contactless screening of a number of subjects in a seamless manner. 


{\it Computer vision methods:}
Computer vision methods are most promising among emerging set of non-contact methods for dehydration monitoring. For example, Liu et. al. utilize smartphone cameras to capture videos/images of the subjects remotely, and subsequently process the frames/images to evaluate hydration-related parameters such as skin elasticity and recovery \cite{sm14}. Another work utilizes a neural network model to analyze face photos, and reports an accuracy of 76.1\% in distinguishing between hydrated and dehydrated states \cite{sm13}. 

\subsection{Discussion}
Non-contact methods are good for mass screening at public places, e.g., stadiums, and during mass events, e.g., marathons, and car rally. They also come handy in the times of pandemics whereby it is important to remotely assess well-being of a person from a distance. 

\section{Multi-Modal Sensing Methods}

Though a majority of the literature on hydration monitoring focuses on a single sensing modality, nevertheless, recently there have been efforts to integrate multiple sensing modalities into a single IoMT device with the aim to overcome the limitations of the individual modalities in order to increase the accuracy and reliability of hydration prediction. 

\indent Firstly, there are a few works that aim to design a wearable solution for hydration monitoring based upon EDA data but they collect PPG data as well. The incorporation of additional PPG data helps keeping in mind the limitation of EDA's sensitivity to cognitive stress and stimuli. Under this framework, Suryadevara et. al. detect dehydration with a mean squared error of 2\% for the prediction of total body water loss \cite{suryadevara2015towards}. In another work, Posada-Quintero et al. simultaneously measure EDA and PPG signals, while the subjects undergo a Stroop test\footnote{Under Stroop test, participants are presented with color words printed in different ink colors and are asked to name the ink color while ignoring the meaning of the word.}. They utilize SVM and report an accuracy of 91.2\% in identifying dehydrated subjects \cite{posada2019mild}. 

In addition, there are set of works that demonstrate superior hydration monitoring performance due to multi-modal sensing, done by different combinations of various kinds of IoMT sensors. For example, Ring et al. show that augmenting bioimpedance data with temperature information significantly improves the inference of water loss estimation of sportsmen \cite{7182268}. In another work, authors obtain bio-impedance and optical spectroscopy measurements, and correlate them with skin hydration levels \cite{22}. Moreover, there are works that combine thermal and electrical properties of skin to detect skin hydration levels \cite{29}. This approach has been tested and validated on porcine skin samples as well as in-vivo tests involving a large number of subjects \cite{27}. 

There have been efforts to design multi-modal biochemistry methods for in-situ urine analysis for hydration assessment as well. Solovei et al. introduce a non-invasive approach for continuous dehydration monitoring by evaluating skin-relative humidity, body weight, and potassium ion concentration in urine using a nanostructured titanium dioxide humidity sensor and a potassium-ion-selective electrode \cite{R_Urine_4}. Wang et al. do hydration status prediction by combining various physiological and sweat biomarkers, including heart rate, core temperature, sweat sodium concentration, and total body sweat rate \cite{48}. 

\subsection{Discussion}
The recent advances in device fabrication, manufacturing and miniaturization have enabled the integration of multiple IoMT sensors into a single compact wearable device. For example, an IoMT device for hydration monitoring has been demonstrated recently that includes an accelerometer, magnetometer, gyroscope, EDA sensor, PPG sensor, temperature sensor, and barometric pressure sensor \cite{17}. Another IoMT device demonstrates the integration of ultrasonic sensors, load cells, and BIA modules in one setup and achieves high accuracy in detecting dehydration \cite{9293804}.
All in all, the multi-modal sensing-based studies highlight the advantages of integrating multiple sensors and multiple biomarkers to enhance hydration level prediction.

\section{Patient-Centric Hydration and Water Assessment in Targeted Body Organs}


In addition to the works that aim to assess the peripheral or internal hydration level of a person, there exists another set of works that aim to quantify the amount of water/fluid accumulated in specific organs of vulnerable population groups under pathological conditions. Such imbalanced water content in some organs can be life-threatening for certain patients, such as in cases of excessive water in the brain \cite{64} or lungs \cite{12}.\\
\indent One such work deals with the patients hospitalized with acute ischemic strokes. Such patients often experience sudden neurological deficits caused by focal cerebral ischemia. Thus, assessing their hydration levels is crucial, as dehydration can exacerbate neurological damage and worsen clinical outcomes. To address this challenge, Bahouth et al. explore the use of a non-invasive cardiac output monitor (NICOM) to assess hydration status, comparing its effectiveness with the BUN/creatinine ratio \cite{66}, which is an indirect indicator of volume depletion. The findings reveals that the NICOM assessment is well-tolerated by 29 out of 30 patients and shows a 70\% agreement with the BUN/creatinine ratio.\\
\indent Another study focuses on hydration monitoring of wounds using a biosensor consisting of a LM35 temperature sensor and a MAX30100 heart rate sensor, for measuring the temperature and blood oxygenation levels, two important indicators of wound healing process \cite{35}.
In another interesting work \cite{65}, Marino et. al. focus on patients with acute heart failure which often require careful monitoring of their hydration levels due to the risk of fluid overload due to impaired cardiac function causing excess fluid accumulation in the body. They conduct bioelectric impedance vector analysis which reveals a reduction in hydration status during the recovery process of acute heart failure patients, indicating correct assessment and the effective removal of accumulated fluids.\\
\indent In short, monitoring abnormal fluid levels in specific organs of the chronic patients accurately is a related problem with high clinical significance. Due to its similarities with the classical hydration level assessment problem, we are of the opinion that many of the techniques in the existing suite of non-invasive and non-contact methods could be repurposed for this problem. Adopting a patient-centric approach and developing efficient wearable or contactless solutions for fluid monitoring in targeted organs will enable real-time, personalized hydration management. This is crucial, as overhydration, dehydration, or even minor imbalances can lead to serious health risks.

\section{Datasets}

Keeping in mind the rise of AI methods and noting that data is the key/backbone for success of AI methods, we have also prepared a list of publicly available and private datasets (available upon request) for hydration assessment (see Table \ref{table:public_datasets}). While making this table, we took care to include key details such as the sensing modalities employed and the devices used for data collection, number of participants, etc. 
Table \ref{table:public_datasets}) also reveals that there do not exist medically annotated public datasets for this problem. That is, there are no such datasets which contain non-invasive or non-contact sensing data of actual dehydrated patients in a hospital setting, along with their blood and urine samples based biochemistry reports. This is what has limited the researchers to study dehydration severity classification problem only. In other words, despite the fact that biochemistry-based biomarkers, e.g., blood/plasma osmolality (in mmol/L), are considered the gold standard for dehydration monitoring \cite{b5}, to date there exists no work in open literature that aims to infer them in non-invasive and non-contact manner. 

\begin{table*}[h!]
\centering
\begin{tabular}{|c|p{1.3cm}|p{6cm}|p{6cm}|p{2cm}|} 
\hline \rowcolor{lightgray}
{\bf Work} & {\bf Availability }& {\bf Device} & {\bf Modality} & {\bf Subjects} \\
\hline

\hline \rowcolor{verylightgray}

\cite{3} & On request. & The developed sensor.& permittivity, conductivity, and resonance frequency.
& In vitro.  \\

\hline
\cite{5} & On request. & The designed electrodes. &Bio-impedance. & 6.   \\

\hline \rowcolor{verylightgray}
\cite{12} & On request. & Electrodes, microphone, temperature sensor, and inertial measurement units. & Bio-impedance, lung sounds, impedance pneumography, temperature, and kinematics.& 24.  \\

\hline
\cite{15} & On request.  & Ultrasonography  GE Logic-E ultrasound machine. & Diameters of the IVC and AO.& 124.   \\

\hline \rowcolor{verylightgray}
\cite{17} & Public. & GSR (Shimmer3). & EDA, PPG accelerometer, gyroscope, temperature, magnetometer, and pressure. & 11. \\



\hline
\cite{alaslani2024you}& Public. & Smartphone. & Video-PPG. & 25. \\

\hline \rowcolor{verylightgray}
\cite{35} & On request.  & LM35 sensor and MAX30100 heart rate sensor. & Body temperature and oxygenation.  & 5. \\

\hline
\cite{57} & On request. & Device developed in \cite{9}. &  Permittivity, weight, and urine specific gravity. & 10.  \\

\hline \rowcolor{verylightgray}
\cite{59} & On request. &  Chest strap heart monitor(the Polar H10 model). &  Heart rate and Orthostatic Changes. & 20. \\

\hline

\cite{73} & Public. &  Epidermal Thermal Sensing Array , OCT, and corneometer (MPA580). &  Thermal properties,  epidermis thickness, stratum corneum thickness, and hydration levels. & 25. \\
\hline \rowcolor{verylightgray}


\cite{22} & On request. & The developed device. & Tetra-polar bio-impedance and optical spectroscopy.& In-vivo\&ex-vivo.   \\
\hline \rowcolor{verylightgray}

\cite{24} & Public. &The designed device.& Thermal properties.&  In-vivo\&in-vitro.\\
\hline

\cite{R_Urine_1} & On request. &Strip. & Urine specific gravity.&  28.\\
\hline \rowcolor{verylightgray}

\cite{ppg2} & On request. & Smartwatch.  & Accelerometer, gyroscope, and PPG signals. &  19.\\
\hline
\end{tabular}
\caption{A quick summary of publicly available datasets for non-invasive dehydration monitoring}
\label{table:public_datasets}
\end{table*}

\section{Discussion}

Table \ref{table:sensing_modalities_comparison} provides a detailed comparison table outlining the benefits and limitations of different sensing modalities for non-invasive and non-contact dehydration monitoring. We observed that each sensing modality observes the body in a different way, and thus, is able to provide unique insights about dehydration pathophysiology. For example, PPG-based methods are good at determining the internal hydration level by measuring the changes in the blood volume. On the other hand, EDA sensing is good at capturing the (peripheral) skin hydration level, while the BIA approach is capable of measuring the intracellular and extracellular hydration levels. Further, the acoustic sensing around the throat could be utilized to measure the daily water intake of a person. Furthermore, a single sensing modality cannot provide full picture about the dehydration level of a person. For example, electrodermal activity sensing could help measure (peripheral) skin hydration only, and not internal hydration level. Also, each sensing modality has its own limitations, e.g., EDA is sensitive to cognitive stress and minor movements, making it less practical in urgent situations where relaxation is difficult. 
Last but not the least, all sensing modalities exhibit a great amount of inter-subject variability, due to factors such as skin tone, skin thickness, aging, temperature. Furthermore, almost all sensing modalities have low sensitivity, i.e., they suffer from the inability to detect subtle/small changes in dehydration level. 

Table \ref{table:existing_surveys} summarizes the contributions and limitations of the existing survey papers on dehydration monitoring. As can be seen, there do exist a few review articles that aim to summarize the state-of-the-art on dehydration monitoring, but these works either put a narrow focus on a particular disease (e.g., Urolithiasis aka kidney stone disease \cite{tahar2024non}), or specific populations, such as elderly \cite{14d}, athletes \cite{70}, etc. Further, none of the existing review articles have talked about contactless dehydration sensing methods. Similarly, none of the existing review articles provides a systematic discussion about the dehydration-centric datasets.

\begin{table*}[h!]
\centering
\setlength{\tabcolsep}{2pt} 
\renewcommand{\arraystretch}{1} 
\footnotesize 
\begin{tabular}{|c|p{6cm}|p{9.5cm}|} 
\hline \rowcolor{lightgray}
 \bf Modality & \bf Pros & \bf Cons
\\
\hline
\bf EDA  & Non-invasive.\newline Can be integrated into daily tools.\newline Achieved a high accuracy. & Susceptibility to external factors such as temperature, sweat, and emotional stress.\newline Vulnerability to postural changes and sudden movements.\\
\hline \rowcolor{verylightgray}
\bf ECG  & Widespread technology.\newline Non-invasive and can be integrated into daily tools. & Primarily dedicated to monitoring heart rate.\newline Limited validation against gold standard methods. \\

 \hline
 \bf Fluid-Based  & Non-invasive and affordable.\newline More suitable for clinical use.\newline Non-invasive. & Not suitable for daily use and requires hygienic precautions.\newline Vulnerable to external factors.\newline Variability in the composition of body fluids among subjects. \\

\hline \rowcolor{verylightgray}
\bf BIA  & Non-invasive and can be integrated into daily tools.\newline Advancement in sensor design. & Susceptibility to external factors such as temperature and sweat.\newline Vulnerability to postural changes and movements. \\

\hline
\bf Acoustic  & Non-invasive and can be contactless.\newline Achieved a high precision by measuring IVC/AO ratio. & Not suitable for daily use and costly.\newline Requires specialized equipment.\newline Requires specific cut-off thresholds tailored, for example to weight. \\

\hline  \rowcolor{verylightgray}
\bf RF & Ability to detect subtle changes in hydration levels.\newline Non-invasive and can be contactless. & Costly and requires specialized equipment.\newline Not suitable for daily use. \\

\hline
\bf Optical  & Can be integrated into daily tools.\newline Non-invasive and can be contactless. & Vulnerable to environmental factors.\newline Initial setup and calibration can be complex.\newline Limited validation against gold standard methods. \\

\hline \rowcolor{verylightgray}
\bf Thermal & Non-invasive.\newline Can detect localized and subtle variations.\newline Can be integrated into daily tools. & Vulnerable to environmental factors like ambient temperature.\newline Initial setup and calibration can be complex.\newline Limited validation against gold standard methods. \\ 

\hline
\end{tabular}
\caption{Comparison of different sensing modalities for dehydration monitoring. }
\label{table:sensing_modalities_comparison}
\end{table*}

\begin{table*}[h!]
\centering
\begin{tabular}{| p{0.7cm}|  p{5.5cm}| p{4cm}| p{6cm}|} 
 \hline \rowcolor{lightgray}
 \bf Work & \bf Scope & \bf Sensing modalities and technologies discussed &  \bf Limitations \\
\hline

\cite{p1} 2022& Iot technologies for active ageing, older people dehydration case & Electrical and Thermal properties &
- The discussion included only a Few sensing modalities.
- No discussion of non-contact techniques.
- Few papers discussed.
- Lack of future perspectives.\\

\hline \rowcolor{verylightgray}
\cite{p2} 2021& Fluid intake monitoring systems for the older people & Thermal, RFID, Inertial Measurement, EMG, Respiratory Inductance Plethysmography, etc & - Assess fluid intake activities instead of hydration as a biological marker.\\
\hline
\cite{p3} 2019& Recent developments in printing flexible and wearable sensing electronics for healthcare applications & BIA, sweat osmolarity, radio-frequency & - Could discuss more sensing modalities.
- Could be organized better.\\
\hline  \rowcolor{verylightgray}
\cite{60} 2017& Methods of assessment of hydration status and their usefulness in detecting dehydration in older people &Body mass, Fluid Based, Neutron activation analysis and BIA. & - Could discuss more sensing modalities. - No discussion of non-contact techniques. - Could discuss more papers. \\
\hline 

\cite{70} 2006& Hydration assessment of athletes &  Body mass, Fluid Based and BIA.& - The discussion included only a Few sensing modalities. - No discussion of non-contact techniques. - Few papers discussed.\\
\hline  \rowcolor{verylightgray}
\cite{p4} 2023&Noninvasive Monitoring to Detect Dehydration: Are We There Yet?&Fluid based, Symptomatology, BIA, ultrasound Based, Nuclear Magnetic Resonance based, electromagnetic based.  &- No discussion of non-contact techniques.\\
\hline 
\cite{p5} 2015&Assessment of hydration status in a large population& Body weight, BIA and fluid based&- The discussion included only a Few sensing modalities. - No discussion of non-contact techniques.\\
\hline \rowcolor{verylightgray}

\cite{p6} 2016&Hydration status assessment techniques and their applicability among Olympic combat sports athletes& BIA and Fluid based&- The discussion included only a Few sensing modalities. - No discussion of non-contact techniques. - Could discuss more papers.\\
\hline
\cite{p7} 2023&Hydration Assessment Methods in the older Population: Advantages and Disadvantages & BIA, weight changes, Fluid based and Symptomatology &- The discussion included only a Few sensing modalities. - No discussion of non-contact techniques. - Few papers discussed.\\
\hline  \rowcolor{verylightgray}

\cite{p8} 2023&Hydration Status Assessment in Older Patients&BIA, Ultrasonography and Fluid based&- Could discuss more sensing modalities. - Few papers discussed. - Could be organized better.
\\
\hline 

\cite{p9} 2019& Use of bedside ultrasound to assess fluid status & ultrasound &- Focus on specific sensing modality. \\
\hline  \rowcolor{verylightgray}

\cite{p10} 2022 &Review of advances in the measurement of skin hydration based on sensing of optical and electrical tissue properties.& BIA and Optical properties & - Focus on specific sensing modality. \\
\hline 

\cite{p11} 2010& A review of the literature on dehydration in the institutionalized older people & Fluid based and weight loss.& - The discussion included only a Few sensing modalities. - No discussion of non-contact techniques. - Could discuss more papers. - Lack of future perspective. \\
\hline  \rowcolor{verylightgray}
\cite{p12} 2015&Clinical and diagnostic utility of saliva as a non-invasive diagnostic fluid & Salivary markers& - Focus on specific sensing modality. - No discussion of non-contact techniques.\\
\hline 

\cite{p13} 2021&The validity of urine color as a hydration biomarker within the general adult population and athletes & Urinary markers& - Focus on specific sensing modality. - No discussion of non-contact techniques.\\

\hline \rowcolor{verylightgray}

\cite{p14} 2018&Evaluation and review of body fluids saliva, sweat, and tear compared to biochemical hydration assessment markers within blood and urine& Fluid based& - Focus on specific sensing modality. - No discussion of non-contact techniques.\\
\hline 
\cite{p15} 2003&Hydration: issues for the 21st century& BIA, body weight, Fluid based& - The discussion included only a Few sensing modalities. - No discussion of non-contact techniques.\\
\hline \rowcolor{verylightgray}
\cite{p16} 2017&A review on opportunities to assess hydration in wireless body area networks& BIA, fluid-based, body mass, 
"Human Body as a Transmission Channel" and RF&- No discussion of non-contact techniques. - Could discuss more papers.\\
\hline 
\cite{p17} 2020&Assessment of hydration status in peritoneal dialysis patients: validity, prognostic value, strengths, and limitations of available techniques& Fluid baaed, BIA and IVC diameter. &    - Could discuss more sensing modalities.\\ 
\hline \rowcolor{verylightgray}
\cite{tahar2024non} 2024& Non-invasive approaches to hydration assessment: a literature review& BIA, EDA, ECG, acoustic signals, PPG, urine color, and motion sensing. &- Could discuss more papers. - No discussion of non-contact techniques. - The paper is dedicated to patients with kidney disease. \\
\hline
\end{tabular}
\caption{Limitations of existing review papers on dehydration monitoring. }
\label{table:existing_surveys}
\end{table*}

\section{Challenges and outlook}

\subsection{Challenges}
Dehydration, being a multi-faceted and whole-body phenomenon, has remained a matter of intense debate among the subject specialists, but without any fruitful conclusion. Thus, there is a lack of consensus among the clinicians and physicians on a universal definition of dehydration. 
In addition, the design of non-invasive and non-contact dehydration monitoring methods is hindered due to a number of challenges as follows. 
First and foremost, a single sensing modality cannot provide full picture about the dehydration level of a person. For example, electrodermal activity sensing could help measure (peripheral) skin hydration only, and not internal hydration level. Secondly, each sensing modality has its own limitations, e.g., EDA is sensitive to cognitive stress and minor movements, making it less practical in urgent situations where relaxation is difficult.  Thirdly, all sensing modalities exhibit a great amount of inter-subject variability, due to factors such as skin tone, skin thickness, aging, temperature. Furthermore, almost all sensing modalities have low sensitivity, i.e., they suffer from the inability to detect subtle/small changes in dehydration level. \\
\indent Another major limitation is that there do not exist medically annotated public datasets for this problem. That is, there are no such datasets which contain non-invasive or non-contact sensing data of actual dehydrated patients in a hospital setting, along with their blood and urine samples based biochemistry reports. This is what has limited the researchers to study dehydration severity classification problem only. In other words, despite the fact that biochemistry-based biomarkers, e.g., blood/plasma osmolality (in mmol/L), are considered the gold standard for dehydration monitoring \cite{b5}, to date there exists no work in open literature that aims to infer them in non-invasive and non-contact manner. Finally, from the viewpoint of clinicians, there is a reluctance to adopt the emerging set of wearable and contactless methods, mainly because of the lack of trust due to lack of clinical trials. Thus, there is a pressing need to redo the dehydration severity assessment problem using one or more of the aforementioned sensing modalities in clinical setting.


\subsection{Outlook}

We foresee that the following mechanisms have the potential to overcome the aforementioned challenges for dehydration assessment.

{\it Novel IoMT-based sensing and instrumentation methods:}
Although many non-invasive and non-contact approaches have shown promise in inferring hydration levels, all the existing sensing modalities often suffer from high noise and low precision or require meticulous precautions for accurate measurements. Thus, there is a need to improve instrumentation and sensor design in order to obtain high-quality data for the AI/ML models. In addition, novel sensing mechanisms for dehydration monitoring are always welcome. For example, in a series of works, Conroy et. al. utilize the Inertial Measurement Unit (IMU) sensors in a smartwatch to track the hand and wrist movements in order to detect daily water intake of kidney stone patients as a proxy for hydration monitoring \cite{Conroy2024,Streeper2023}.

\begin{figure}[h!]
\centering
\includegraphics[scale=0.23]{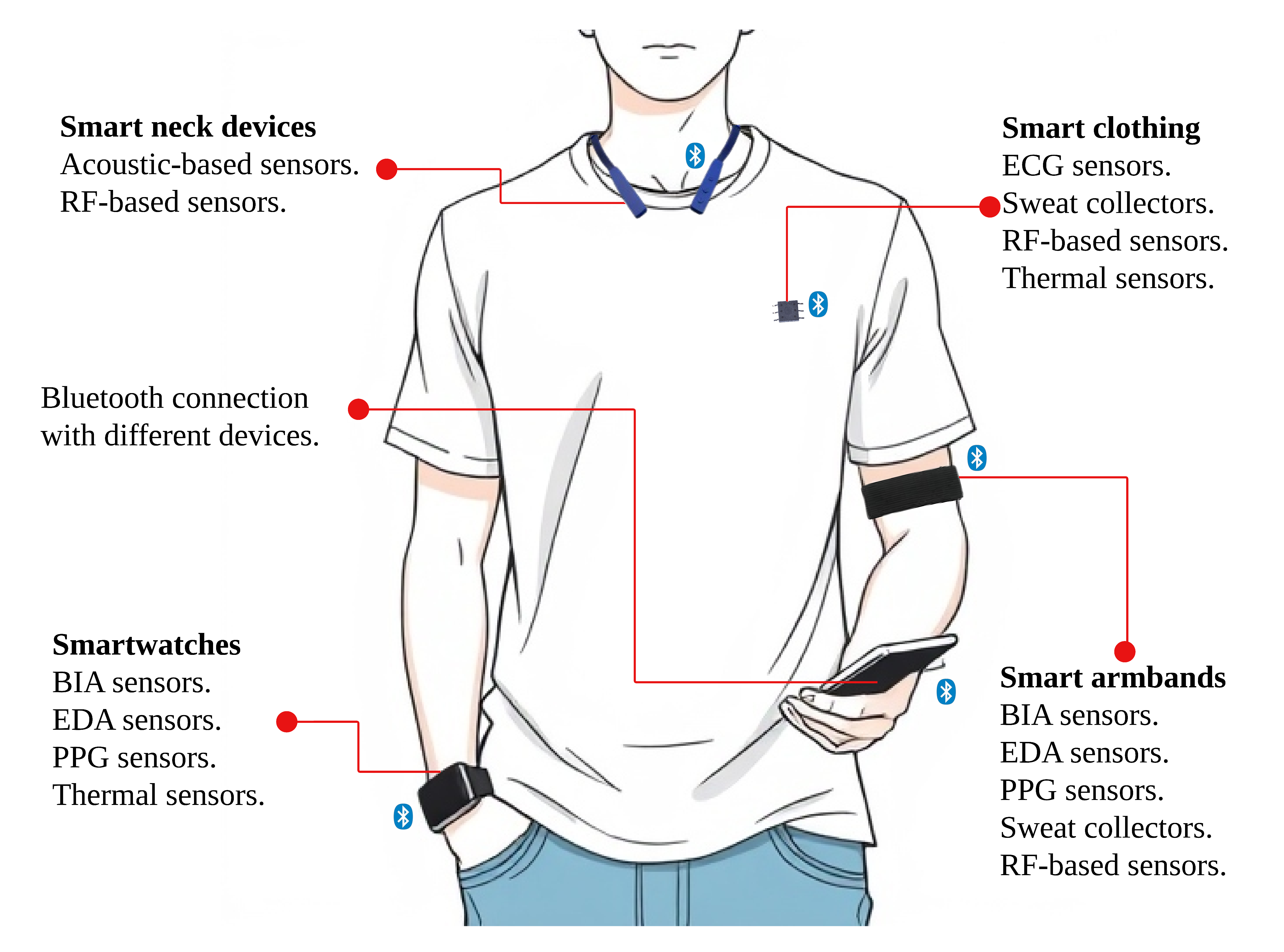}
\caption{Outlook of dehydration monitoring problem: We envision a multi-modal sensing solution for in-situ, seamless, and in-depth monitoring of dehydration level. This diagram provides an illustration of non-invasive dehydration monitoring in near future using various IoMT (Internet of Medical Things) devices equipped with multiple sensing modalities for hydration monitoring, including smart neck devices, smart clothing, smartwatches, and smart armbands. The figure highlights the integration of sensors such as BIA, EDA, PPG, ECG, RF-based, acoustic-based, thermal sensors, and sweat collectors, along with Bluetooth connectivity for seamless data transmission between devices. Here, EDA sensor measures skin hydration, BIA sensor measures the inter-cellular and intra-cellular water, PPG measures blood volume, while acoustic sensor on the neck monitors the daily water intake. }
\label{multi}
\end{figure}

{\it Use of multi-modal sensing and wearables:}
Even after we are able to improve instrumentaion of IoMT devices to obtain high-precision and high quality sensor data, another challenge still persists. That is, each individual sensing modality could provide only limited inference about the hydration level of a person. Thus, it is need of the hour to investigate the design of novel multi-modal AI/ML methods that aim to utilize comprehensive multi-parametric data that comes from an integrated suite of multiple sensors. Such multi-modal sensing-based, and edge-AI-empowered wearable devices could help integrate hydration monitoring into our daily lives to ensure the body's internal balance and normal functioning.
Figure \ref{multi} demonstrates our vision for the future of dehydration monitoring research which lies in multi-modal sensing. Figure \ref{multi} shows that IoMT devices such as smartwatches, smart clothing, and smart neck devices which are equipped with a number of sensors, including BIA, EDA, PPG, ECG, RF-based sensors, thermal sensors, and sweat collectors could comprehensively monitor the dehydration level of a person. Further, these devices can seamlessly interconnect via Bluetooth, enabling real-time data collection. The information collected can be analyzed and assessed using mobile applications, providing users with instant feedback on their hydration state.

{\it Role of ML/AI to cater for inter-subject variability:}
Recall that assessment of body hydration level through non-invasive and non-contact methods involves measuring various markers as a proxy for dehydration, e.g., body/skin conductivity, sudden changes in body mass, total body water, electrolyte proportions in the blood, permittivity, and more. However, factors like age, sex, weight, body temperature, skin thickness, skin tone, skin color... etc., can significantly affect these markers, making it challenging to accurately determine hydration levels. 
One way to eliminate or alleviate the inter-subject variability problem is to calibrate and fine-tune the AI/ML models exclusively for each subject. The alternate solution involves a massive data collection campaign to collect enough data from people with different ethnicity, different skin tones, different work styles (indoors vs. outdoors)... etc., in order to train AI/ML models that are generalizable and robust.


{\it Role of generative AI for generation of synthetic dehydration datasets:}
As mentioned before, the dehydration problem, by definition, is a hard problem, and AI/ML based solutions rely upon finding the relationships between the raw data (that is measured in non-invasive or non-contact manner) and the ground truth (biochemistry report-based biomarkers derived from blood and urine samples). However, the labeled data for dehydration severity classification problem--where AI/ML models predict whether a person is mildly dehydrated, or moderately dehydrated, or severely dehydrated--is very scarce with very few public datasets, while the labeled data for dehydration regression problem--where one aims to infer, say, the plasma osmolality--is non-existent, mainly due to peculiar nature of the data, due to delays and obstacles in getting necessary approvals. Thus, there is an urgent needs to collect medically annotated datasets, i.e., raw data along with biochemistry reports of blood and urine samples. Once such datasets become available, generative AI methods, e.g., diffusion models, Generative Adversarial Networks (GAN), Variational Autoencoders (VAE), could step in, to produce additional high-quality synthetic data to aid the AI/ML regression models in their task to infer plasma and urine osmolality.  

{\it Need for custom solutions for the vulnerable groups:}
There is a dire need to develop dehydration detection and classification methods that are tailored for specific population groups.
As an example, \cite{R9} reports the results of dehydration monitoring in soccer players. They conclude that body mass change is the most reliable indicator of sudden changes in hydration levels, which outperforms other metrics such as hematocrit, USG, and color. Such methods could prove to be useful to athletes as well who lose a lot of water during intense physical activity can greatly benefit from hydration monitoring. Additional such investigations are needed for other vulnerable population groups, e.g., elderly, newborns, people with chronic diseases. 
In-situ dehydration monitoring of elderly people is needed because they may not always feel thirsty but still need adequate hydration. Parents may also need to monitor the hydration levels of infants and newborns. Furthermore, chronic kidney disease patients need to be cautious about the balance of fluids in their body, as over-hydration and dehydration can be life-threatening for them.

{\it Design of more fine-grained dehydration assessment methods:}
It is need of the hour to develop novel hydration monitoring methods for more fine-grained dehydration assessment. For example, it is essential to design new methods that could automatically differentiate between different dehydration types, and the root cause. In addition, it is important to select most appropriate dehydration biomarker for a given patient/subject. For instance, blood sample-based biomarkers might be more suitable for chronic fluid deficits in the elderly individuals, while urine measurements could detect sudden fluid changes \cite{60}. Additionally, specific cutoff values and thresholds based on age, height, and weight are needed for some sensing modalities, such as acoustic techniques, to improve accuracy. Testing on a diverse population with varying body types and ages is necessary.
Last but not the least, measuring hydration level of a EU-hydrated person is a major challenge that requires further attention.

{\it Need for clinical trials:}
A majority of the efforts on design of wearable and contactless methods for dehydration monitoring have been led by the researchers with engineering background. But, from the viewpoint of clinicians, there is a reluctance to adopt the emerging set of wearable and contactless methods, mainly because of the lack of trust due to lack of clinical trials. Thus, there is a pressing need to redo the dehydration severity assessment problem using one or more of the aforementioned sensing modalities in clinical setting.


\section{Conclusion}
Dehydration could cause a significant decrease in total body water, impair the water-electrolyte balance of human body, and could have very fetal implications, e.g., hypotension, seizures, fits, and even death. 
This article critically examined a wide range of emerging non-invasive, and non-contact methods for dehydration monitoring, and summarized the benefits and limitation of each sensing modality. We observed that each sensing modality observes the body in a different way, and thus, is able to provide unique insights about dehydration pathophysiology. For example, PPG-based methods are good at determining the internal hydration level by measuring the changes in the blood volume. On the other hand, EDA sensing is good at capturing the (peripheral) skin hydration level, while the BIA approach is capable of measuring the intracellular and extracellular hydration levels. Further, the acoustic sensing around the throat could be utilized to measure the daily water intake of a person. Thus, to pave the way for the healthcare systems of tomorrow, it is imperative to design novel edge-AI-empowered wearable devices that utilize multi-modal approach to provide a comprehensive understanding of dehydration level of a person.\\
\indent We also observed a lack of medically annotated public datasets for this problem. That is, to the best of our knowledge, there doesn't exist datasets where raw sensing data is collected from dehydrated patients admitted to a hospital, along with their blood and urine samples based biochemistry reports. Another important problem that deserves the attention of the community is to develop non-invasive and non-contact methods to automatically identify the type of dehydration, and the root cause. Last but not the least, measuring hydration level of a EU-hydrated person is a major challenge that requires further attention. Nevertheless, we believe it is high time to make hydration assessment a daily practice and integrate it into smartphones, wearable devices, and smartwatches to realize patient-centric healthcare systems of future, which are in turn a key ingredient of smart homes and smart cities of future. 


\section*{Acknowledgements}
Ana Bigio, a scientific illustrator, produced Figure 2. 

\footnotesize{
\bibliographystyle{IEEEtran}
\bibliography{references}
}




\vfill\break

\end{document}